\documentstyle[11pt,appbx,epsfig]{article}
\begin{document}
\title{Reinvestigation of $pp\rightarrow pp\pi^0$ and
       $pp\rightarrow pp\eta$ at threshold
\thanks{Presented at the "Meson 98" Workshop, Cracow, Poland, May 29 -- June 2,
1998, Preprint-No.\ FAU-TP3-98/9 (to be published in ACTA PHYSICA POLONICA B)}}
\author{Frieder KLEEFELD and Manfred DILLIG
\address{
Inst.\ f.\ Theoretical Physics III, University of Erlangen-N\"urnberg,
Staudtstr. 7, 91058 Erlangen, Germany\\
kleefeld@theorie3.physik.uni-erlangen.de,
mdillig@theorie3.physik.uni-erlangen.de}
}
\maketitle

\begin{abstract}
$pp\rightarrow pp\pi^0$ and $pp\rightarrow pp\eta$ were considered close to
threshold. The pion production was found not only to be dominated by
IA and rescattering contributions, but also strongly by box-diagrams, resonances and
heavy-meson exchanges. New
relevant contributions for the IA were found.
The role of the offshell behaviour of nucleon-resonance
couplings ($P_{11}(1440), S_{11}(1535)$) and of interference effects for the eta production was 
investigated. The results opened a large door for microsopic understanding of short
range physics. A covariant model for $pp\rightarrow pp\eta$ being under construction is going to be
finished.
\end{abstract}

\section{Reinvestigation of $pp\rightarrow pp\pi^0$ at threshold}

\begin{figure}[hp]
\epsfxsize=  13.0cm
\epsfysize=  8.0cm
\centerline{\epsffile{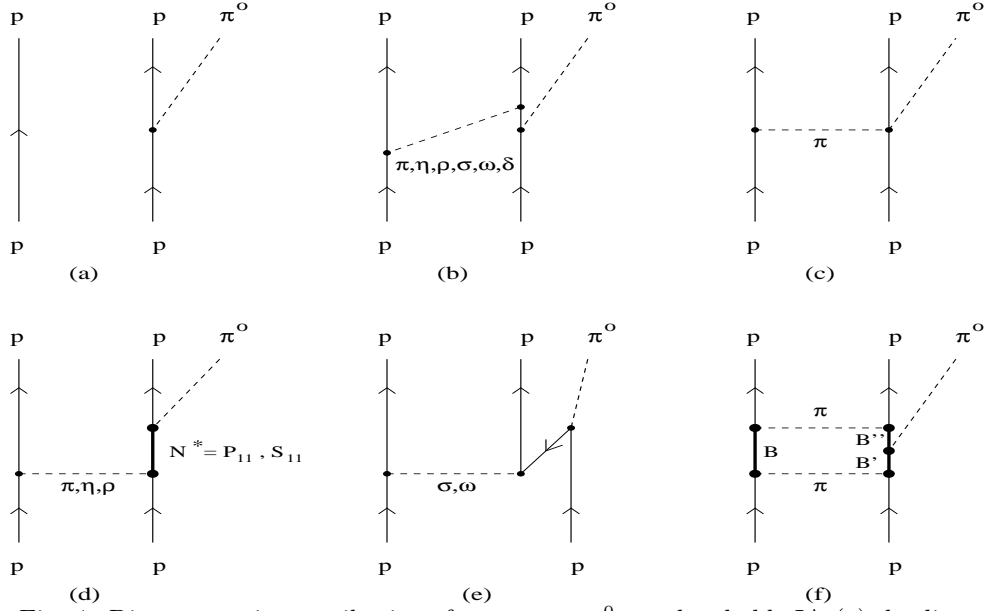}}
\caption{Diagrammatic contributions for  
$pp\rightarrow pp\,\pi^0\;$ at threshold: 
IA (a), leading irreducible correction to IA (b), non-resonant 
S--wave-rescattering (c), resonant S-- and P--wave contributions (d), 
Heavy meson exchange (e), $\pi\pi$ box-diagrams (f).} \label{abbpi1}
\end{figure}

\begin{figure}[hp]
\epsfxsize=  13.0cm
\epsfysize=  8.0cm
\centerline{\epsffile{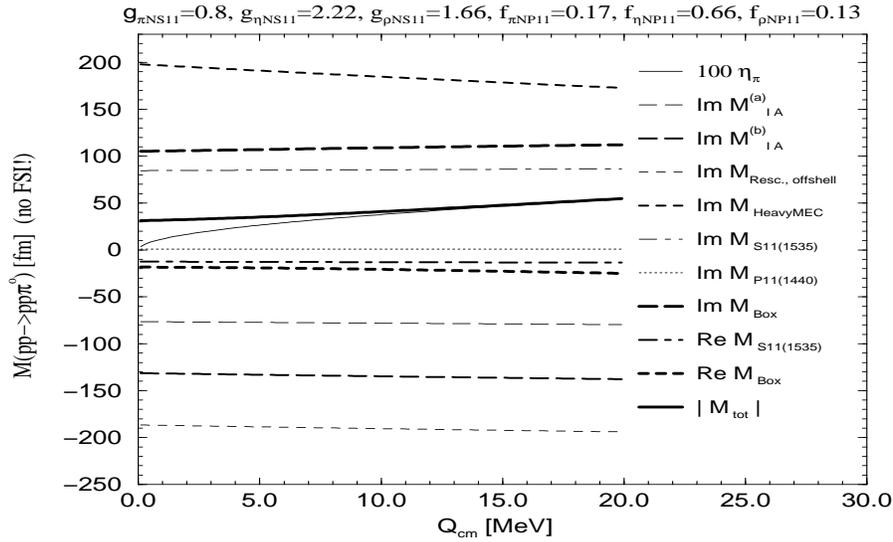}}
\caption{Nonvanishing real-- and imaginary parts of $ 
<\!{\cal
M} \, (Q_{cm})\!>$.}
\label{mmmxy0}
\end{figure}

Following Watson-Migdal we multiplicatively can separate the shortranged
production part of the matrix elements from the long-ranged part 
belonging to (initial and) final state interactions (ISI, FSI). Under the assumtion that
the short range part approximately is a slowly variing function of the 
phasespace integration variables  
the total production cross section of $pp\rightarrow pp\pi^0$ at threshold can be calculated in
the following way: 

\begin{eqnarray} \lefteqn{\sigma^{\, \mbox{\scriptsize SFSA}}_{pp\rightarrow pp\,\pi^0} (s)
 \quad =  
\frac{1}{2 \, \sqrt{\lambda (s,m^2_p,m^2_p)}} \; 
 \frac{1}{(2\,\pi )^5} \; 
\; 
 R^{\, \scriptsize \mbox{FSI}}_3 (s \; ; \; m^2_p, m^2_p, m^2_{\pi^0} )
\; 
  \cdot }
 \nonumber \\
 & & \nonumber \\
 & \cdot & 
 \Big| 
<00,11| \; \Big( \;
{\cal M}_{\, \mbox{\scriptsize IA}}^{\, \mbox{\scriptsize (a)}} (\vec{p}\,) \,
+
{\cal M}_{\, \mbox{\scriptsize IA}}^{\, \mbox{\scriptsize (b)}} (\vec{p}\,) \,
+
{\cal M}_{\, \mbox{Rescatt.}} (\vec{p}\,) \,
+
{\cal M}_{\, S_{11}} (\vec{p}\,) \,
+
{\cal M}_{\, P_{11}} (\vec{p}\,) 
\nonumber \\
 & & \nonumber \\
 & & + \;
{\cal M}^{HMEC}_{\, \sigma + \omega} (\vec{p}\,) \,
+ 
\sum\limits_{B^{\prime\prime}B^{\prime}B} \; \left[ \;
{\cal M}^{NC}_{\, B^{\prime\prime}B^{\prime}, \, B} \;
(\vec{p}\,) 
+ 
{\cal M}^{C}_{\, B^{\prime\prime}B^{\prime}, \, B} \;
(\vec{p}\,) 
\right] \; 
\Big) \; |10,11>
 \, \Big|^2 \nonumber 
\end{eqnarray}

while  
$R^{\, \scriptsize \mbox{FSI}}_3 (s)$ is a FSI-modified three body phasespace
integral and $s$ the square of the CM-energy
($\; 2 \, |\vec{p}\,| = \sqrt{s\, -\, 4 \, m^2_p}\; $; $B,B',B''=N,\Delta$,
C="Crossed Box", NC="Noncrossed Box"). The different contributions are shown in
Fig.\ \ref{abbpi1}. The resulting matrix elements are shown in Fig.\
\ref{mmmxy0} ($Q_{cm}=$ excess energy)(For couplings and cutoffs used see [1],
p.\ 56 !). Noticable is the large role of the (usually neglected) 
irreducible IA contribution (b) with one exchange pion present when the produced
pion is radiated off. The other remarkable point is the opposite sign of the 
heavy meson exchanges compared to previous
publications ($<{\cal M}_{\sigma}>\, \simeq -208.7$ fm and $<{\cal M}_{
\omega}>\, \simeq 406.7$ fm at threshold), and the nonvanishing contribution of the box-diagrams. 
Using FSI-corrections due to the Reid-Soft-Core- and the Coulomb-potential the
resulting total cross section is about a factor 0.042 smaller then the experiment
at $\eta_\pi =0.31$, which is mostly due to the large HMEC contribution.
It should be noted, that from their branching ratios the following onshell
meson-nucleon-resonance couplings were calculated:

\begin{eqnarray} & & |f_{\pi NP_{11}}| = 0.17 \; , \quad 
|f_{\eta NP_{11}}| = 0.66 \; , \quad
|f_{\rho NP_{11}}| = 0.13 \; , 
 \nonumber \\
 & &
|g_{\pi NS_{11}}| = 0.77 \; , \quad 
|g_{\eta NS_{11}}| = 2.08 \; , \quad
|g_{\rho NS_{11}}| = 0.22 \nonumber
\end{eqnarray}

while for the calculation above we use the couplings used by Moalem et al.\
[2] listed at the Figures. Using the offshell couplings of the following 
$\eta$-production model we mention, that all the above 
$S_{11}(1535)$ resonance contributions to
$\pi^0$-production will vanish! As a result we state that the complex production
mechanism of $pp\rightarrow pp\,\pi^0$ is still not clearly understood.

\section{Reinvestigation of $pp\rightarrow pp\eta$ at threshold}

It is well known for the proton induced $\eta$-production process at threshold that in resonant
meson-exchange models the interference and the relative sign between the different 
meson-exchange amplitudes controls the absolute magnitude of the total
cross-section compared to --- by isospin invariance connected --- processes like
$pn\rightarrow pn\eta$ and $pn\rightarrow d\eta$. By now people fix the 
relative signs of amplitudes by hand in order to obtain and understand the
nature of the total cross section.
To avoid this arbitrariness we tried to calculate the real parts of the relevant
meson-nucleon-resonance couplings 
$g_{\lambda NN^\ast} (\vec{p}^{\;\prime},\vec{p} \, )$ by a model, in which the mesons couple to
quarks in the proton and the respecting resonances (which are considered to be
Harmonic Oscillators described by one unique range parameter $a$. The mixing
angle between spin 1/2 and 3/2 components in the $S_{11}(1535)$-resonance is denoted
by $\theta$, the polarization vector of the exchanged $\rho$-meson by
$\vec{\varepsilon}_\rho$). The resulting couplings are ($g_{\lambda NN^\ast}
(\vec{0},\vec{p} \, )$):
\begin{figure}[h]
\epsfxsize=  14.0cm
\epsfysize=  4.0cm
\centerline{\epsffile{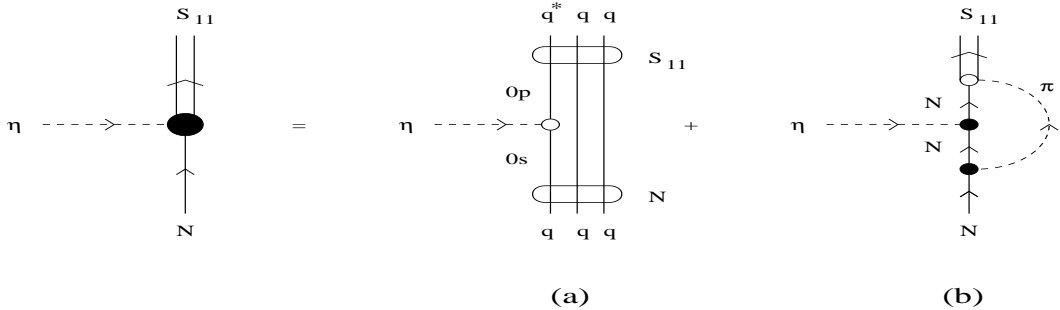}}
\caption{Relevant contributions to the $\eta NS_{11}$--coupling at the
$\eta$--threshold. \label{fff4}} 
\end{figure}

\begin{eqnarray}
g^{(a)}_{\pi NS_{11}} & = &  
 \frac{\sqrt{2}}{3}\,\;
\frac{f_{\pi qq}}{m_{\pi}} \;\;
\;   
\cos\left(\theta -\arcsin\frac{1}{3}\right) \; 
\frac{|\vec{p}\,|^2}{a}\;
 \;\exp (-\,\frac{|\vec{p}\;|^2}{6a^2}) \nonumber \\
 g^{(a)}_{\eta NS_{11}} & = &  
 \frac{\sqrt{2}}{3}\,\;
\frac{f_{\eta qq}}{m_{\eta}} \;\;
\;   
\cos\left(\theta -\arcsin\frac{1}{3}\right) \; 
\frac{|\vec{p}\,|^2}{a}\;
 \;\exp (-\,\frac{|\vec{p}\,|^2}{6a^2}) 
\nonumber \\
 g^{(a)}_{\rho NS_{11}} & = &
 - \, \frac{1}{3}\;\sqrt{\frac{11}{6}}\;
\frac{f_{\rho qq}}{m_{\rho}} \;  
 \cos\left( \theta \, - \, \arcsin \frac{1}{\sqrt{33}}\; \right) \;
\;\frac{|\vec{p}\,|^2}{a}
 \;\exp (-\,\frac{|\vec{p}\,|^2}{6a^2}) \; \left( 1-\frac{(\vec{\varepsilon}_\rho
 \cdot \vec{p} \, )^2}{|\vec{p}\,|^2} \right) \nonumber 
\end{eqnarray} 

\begin{figure}[hp]
\epsfxsize=  14.0cm
\epsfysize=  8.0cm
\centerline{\epsffile{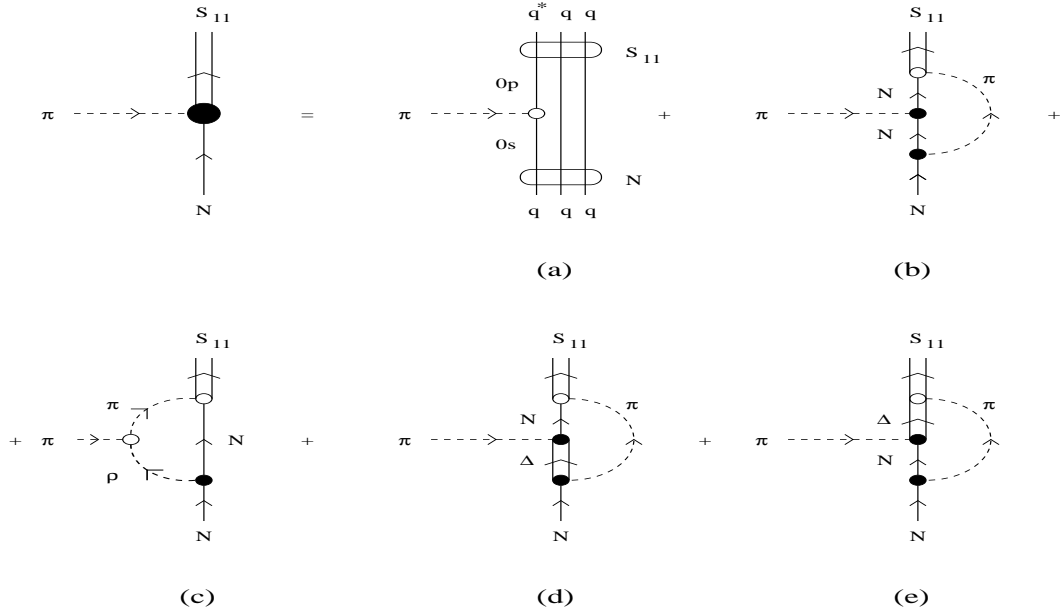}}
\caption{Relevant contributions to the $\pi NS_{11}$--coupling
 at the $\pi$--threshold. \label{fff3}} 
\end{figure}

\begin{figure}[hp]
\epsfxsize=  14.0cm
\epsfysize=  8.0cm
\centerline{\epsffile{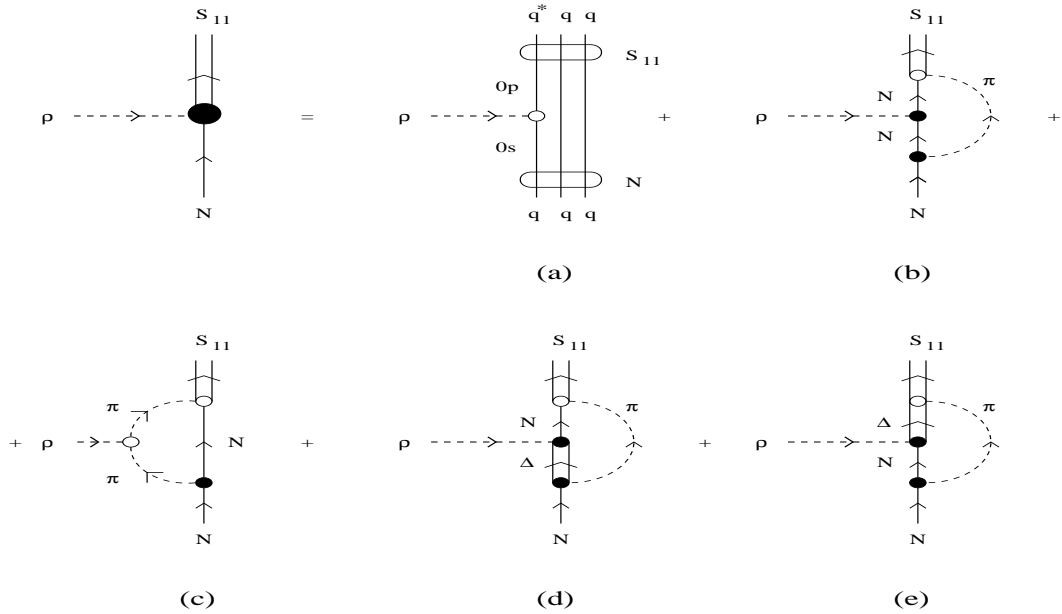}}
\caption{Relevant contributions to the $\rho NS_{11}$--coupling at 
$\rho$--threshold. \label{fff2}} 
\end{figure}

The
imaginary parts of the couplings we calculate by nonrelativistic lowest order 
meson loop corrections to the vertices considered. The considered diagrams are
shown in Figs.\ \ref{fff4}, \ref{fff3}, \ref{fff2} for the $S_{11} (1535)$
resonance. The Roper resonance can be treated analogously. 
Obviously the real part of the couplings is still not fixed, but seems to be
dominated by the quark contributions.
The couplings strongly depend on the mixing angle $\theta$ which can be measured
by this method, if the model takes care for the complete real part of the couplings.
At the moment the value $\theta=-5^0$ is favoured compared to larger values. The
range parameter $a$ shows up to be about $(0.5 \; \mbox{fm})^{-1}$.
As a general result the observations of Moalem et al.\ [2] are confirmed, but the
mechanisms which lead to the absolute magnitude of the total cross section
are very nontrivial. The couplings at the vertex of the produced $\eta$-meson
are in general very small, but the coupling constants at the internal vertices
of the exchanged meson show up to be very large and to have imaginary part!
Putting together all these effects the results of Moalem on average are
obtained. A detailed discussion will be given in [3].\\
\mbox{} \hrulefill \mbox{} \\
\makebox[4mm]{[1]} R.\ Machleidt et al., 
Phys.\ Rep. {\bf 149} (1987) 1 \\
\makebox[4mm]{[2]} A.\ Moalem et al., hep--ph/9505264;
Nucl.\ Phys. {\bf A 589} (1995) 649; \\
\makebox[4mm]{} Nucl.\ Phys. {\bf A 600} (1996) 445 \\ 
\makebox[4mm]{[3]} F.\ Kleefeld, Doctoral Thesis (Univ.\
Erlangen-N\"urnberg, Germany) 

\end{document}